\documentclass[5p]{elsarticle}
\usepackage{amsmath,amssymb,bm,mathrsfs}
\usepackage{srcltx}
\usepackage{dsfont}
\usepackage{epsfig}
\usepackage{slashed}
\usepackage{bbold}
\usepackage{psfrag}
\usepackage{xcolor}
\PassOptionsToPackage{caption=false}{subfig}
\usepackage{subfig,bm}
\usepackage{xfrac}
\usepackage{multirow}
\usepackage{booktabs}
\usepackage[colorlinks=true,linkcolor=blue,citecolor=blue,urlcolor=violet]{hyperref}

\newcommand{\hhref}[1]{\href{http://arxiv.org/abs/#1}{arXiv:#1}}

\newcommand{\be}{\begin{equation}}
\newcommand{\ee}{\end{equation}}
\newcommand{\bea}{\begin{eqnarray}}
\newcommand{\eea}{\end{eqnarray}}

\newcommand{\eq}[1]{eq.~(\ref{#1})}

\biboptions{numbers,sort&compress}

\def\({\left(}
\def\){\right)}

\usepackage{soul}

\begin{document}

\begin{frontmatter}

\title{Diboson Interference Resurrection\vspace{0.4cm}} 

\author[ifae]{Giuliano Panico}
\ead{gpanico@ifae.es}

\author[cern]{Francesco Riva}
\ead{francesco.riva@cern.ch}

\author[cern,epfl,padova,padovaI]{Andrea Wulzer}
\ead{andrea.wulzer@pd.infn.it}

\address[ifae]{IFAE, Universitat Aut\`onoma de Barcelona, E-08193 Bellaterra, Barcelona, Spain} 
\address[cern]{Theoretical Physics Department, CERN, Geneva, Switzerland}
\address[epfl]{Institut de Th\'eorie des Ph\'enomenes Physiques, EPFL, Lausanne, Switzerland}
\address[padova]{Dipartimento di Fisica e Astronomia, Universit\'a di Padova, Italy}
\address[padovaI]{INFN, Sezione di Padova, via Marzolo 8, I-35131 Padova, Italy}

\begin{abstract}
High-energy diboson processes at the LHC are potentially powerful indirect probes of heavy new physics, whose effects can be encapsulated in higher-dimensional operators or in modified Standard Model couplings. An obstruction however comes from the fact that leading new physics effects often emerge in diboson helicity amplitudes that are anomalously small in the Standard Model. As such, the formally leading Standard Model/New Physics interference contribution cancels in inclusive measurements. This paper describes a solution to this problem.

{\vspace{.5em}
Preprint: CERN-TH-2017-185 }
\end{abstract}

\end{frontmatter}

\section{Introduction}
Precision tests are an increasingly important tool to search for dynamics beyond the Standard Model (SM). Effects of heavy new physics on SM processes are captured by an effective field theory (EFT)~\cite{Grzadkowski:2010es}, generically dominated by dimension-6 operators (collectively denoted as ``BSM effects'' in what follows). Some of these induce energy-growth in electro-weak (or even strong, see \cite{Dixon:1993xd,Krauss:2016ely,Alioli:2017jdo}) scattering processes, becoming sensitive targets of the LHC large kinematic reach, provided accurate enough measurements are possible in the high-energy regime. The power of this interplay between energy and accuracy has been demonstrated in ref.~\cite{Farina:2016rws} for the neutral and charged Drell-Yan processes. High-energy diboson \cite{Hagiwara:1986vm,Falkowski:2015jaa,Frye:2015rba,Butter:2016cvz,Green:2016trm,Zhang:2016zsp,Baglio:2017bfe} and boson-plus-Higgs \cite{Ellis:2014dva,Biekoetter:2014jwa} production processes are promising channels to be explored in this context, and they are sensitive to a wider variety of BSM effects than Drell-Yan. A possible obstruction to this program, outlined in ref.~\cite{Azatov:2016sqh}, takes the form of a ``non-interference theorem'', formulated as follows. In the high-energy limit $E\gg m_W$ amplitudes can be well characterized by the helicities of the external bosons. In this regime, $2\to2$ tree-level amplitudes involving  transversely polarized vector-bosons, turn out to exhibit different helicity in the SM and BSM. For instance, the diboson processes that we consider here, $ f f \to W_T V_T$, $V=W,Z,\gamma$ have final-state helicity $(\pm\mp)$ in the SM  and $(\pm\pm)$ in BSM. This implies that SM and BSM do not interfere in inclusive analyses, so that the first departure from the SM appears at order BSM-squared: an important obstacle to a precision program aiming at measuring small effects. In this article we discuss a strategy to ``resurrect'' the SM-BSM interference, based on the measurement of the bosons azimuthal decay angles. Similar measurements were proposed long ago in refs.~\cite{Duncan:1985ij,Hagiwara:1989mx}.\\

\noindent{\bf Note added:} While this work was in preparation \cite{GPFRtalk}, ref.~\cite{Azatov:2017kzw} appeared, that discusses similar ideas in the context of the $WZ$ process.

\section{Interference Resurrection}\label{sec:IR}
We consider the production of two massive vector bo\-sons $V^{1,2}=\{W,Z\}$, followed by fermionic decays $V^{1 (2)}\rightarrow f^{1(2)}_+ f^{1(2)}_-$. The final state fermions are labeled by their helicities, with ``$f$'' denoting irrespectively particles or anti-particles of any charge or flavor. We are mostly interested in $2\rightarrow2$ quark-initiated production, however most of what follows holds for generic diboson production, possibly in association with QCD jets. 

\begin{figure}[ht]
\includegraphics[width=.5\textwidth]{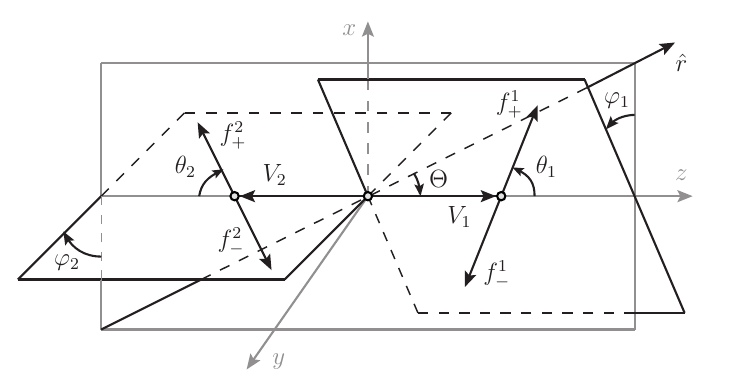}
\caption{Definition of the decay angles for the diboson system.}\label{fig:decay_angles}
\end{figure}

We choose a ``special'' coordinate system, defined as follows. Starting from the lab frame, and a generic configuration for the external state momenta, we first boost back to the center of mass frame of the diboson (or $4$-fermions) system. The boost is as customary performed along the direction of motion (call it $\hat{r}$) of the diboson system. In the new system we have back-to-back boson momenta and  the reference unit vector $\hat{r}$, which we use to define the special frame as shown in fig.~\ref{fig:decay_angles}. Namely, we take the $z$ axis of the special frame along the direction of motion of the first boson $V^1$ while the $x$ axis is in the plane formed by $\hat{r}$ and the diboson axis. The $x$ orientation is taken such that $\hat{r}$ goes in the positive $x$ direction or, equivalently, such that the $y$ axis (for left-handed orientation of the $x$-$y$-$z$ system) is parallel to the cross-product between the $V^1$ direction and~$\hat{r}$. For a $2\rightarrow2$ production process, $\hat{r}$ coincides with the collision axis, oriented in the direction of the parton that carried the larger energy in the lab frame. In the special frame the collision thus occurs in a rather special configuration, where the initial states move in the $x$-$z$ plane while the intermediate bosons happen to be produced exactly parallel to the $z$-axis. 

The reader might be confused by the fact that the special reference system depends on the kinematical configuration of the event, i.e. different systems are employed for the calculation of the amplitude at different phase-space points. The amplitude obtained in this way does not indeed coincide with the one evaluated directly in the lab frame. To obtain the latter out of the former one has to act with the phase-space dependent Lorentz transformation that connects the special frame with the lab, introducing in this way an additional and complicated dependence on the kinematical variables. However the physical external states of the process are the massless helicity eigenstate fermions, and Lorentz transformations act as multiplicative phase factors on massless states helicity amplitudes. Therefore this additional dependence on the kinematics drops from the amplitude modulus square and is unobservable. Stated differently,  the amplitude for each kinematical configuration corresponds to one individual quantum-mechanically distinguishable process. As such, each one can be safely computed in its own frame.

In the special frame the amplitude reads
\begin{equation}\label{amp}
{\mathcal{A}}\propto g_1 g_2 \sum_{h_{1,2}}{\mathcal{A}}_{h_1h_2}e^{i  h_1\varphi_1}e^{i  h_2\varphi_2}d_{h_1}(\theta_1)d_{h_2}(\theta_2)\,,
\end{equation}
where $g_{1(2)}$ are the couplings responsible for the $V^{1(2)}$ decays and ${\mathcal{A}}_{h_1h_2}$ denotes the amplitude for the production of on-shell vector bosons with helicities $h_1$ and $h_2$, evaluated  in the special frame. Normalizations and $\varphi_{1,2}$-dependent overall phases, that will drop from the amplitude modulus square, have been absorbed in the proportionality factor. The above equation relies on the narrow-width approximation for the decaying bosons only to the extent to which it ignores possible Feynman diagrams where the fermion pairs do not originate from the virtual vector bosons, and by the fact that the ``hard'' amplitude ${\mathcal{A}}_{h_1h_2}$ is computed with exactly on-shell bosons. Its validity does not require the fermion pairs invariant masses being exactly equal to the pole mass of the corresponding bosons, though the amplitude is  peaked around this configuration, because of the usual Breit-Wigner factors that we reabsorbed in the normalization factor. 

The variables $\theta_{1(2)}\in[0,\pi]$ are the polar decay angles of each boson in its rest frame, oriented in the direction that goes from the $3$-momentum of the $V^{1(2)}$ boson to the one of the right-handed fermion $f^{1(2)}_+$ produced in its decay. In the special frame they are obtained from the rapidities $\eta$ of the final state fermions by the relations
\begin{eqnarray}\label{theta}
&&\cos\theta_{1}=\tanh\frac{\eta^s(f_+^1)-\eta^s(f_-^1)}2\,,\nonumber\\
&&\cos\theta_{2}=\tanh\frac{\eta^s(f_-^2)-\eta^s(f_+^2)}2\,,
\end{eqnarray}
where the ``$\,^s\,$'' subscript denotes spacial frame quantities. The azimuthal variables $\varphi_{1(2)}\in[0,2\pi]$ are defined in the center of mass frame of the diboson system (see fig.~\ref{fig:decay_angles}) as the angles between the decay plane of each boson and the $x$-$z$ plane of the special coordinate system. The orientation of the decay plane is taken in the direction that goes from $V^{1(2)}$ to $f^{1(2)}_+$. In the special frame, $\varphi_{1(2)}$ are simply the azimuthal angles $\phi$ of the final state fermions. More precisely
\begin{eqnarray}\label{phi}
&&\varphi_{1}=\phi^s(f^1_+)=\phi^s(f^1_-)+\pi\,,\nonumber\\
&&\varphi_{2}=\pi-\phi^s(f^2_+)=-\phi^s(f^2_-)\,,
\end{eqnarray}
modulo $2\pi$. Notice that our seemingly asymmetric definition of the decay angles for the two bosons is actually what is needed to describe their decay symmetrically in their own rest frames. Indeed it produces $1\leftrightarrow2$ symmetrical angular factors in eq.~(\ref{amp}).

With these definitions, eq.~(\ref{amp}) is easily obtained by direct calculation or by applying the Jacob--Wick partial wave decomposition formula~\cite{Jacob:1959at} to the case of a $J=1$, $m=h$ particle decaying to two particles with helicity difference $\lambda=\lambda_1-\lambda_2=+1$.\footnote{The result does depend on conventions in the definition of the vector boson polarization vectors: different definitions can produce  phases in the vector boson decay amplitudes, that compensate for the extra phases that will emerge from the diboson amplitude calculation. The standard HELAS conventions~\cite{Murayama:1992gi} are employed here.} Partial wave decomposition determines the $\varphi_{1(2)}$-dependent phase factors in eq.~(\ref{amp}) (up to the previously mentioned overall phases) and gives us $d_h(\theta)$ equal to the $d_{m,\lambda}^{J}$ Wigner function, i.e.
\begin{equation}
d_{\pm1}(\theta)=\frac{1\pm\cos\theta}2\,,\;\;d_{0}(\theta)=\frac{\sin\theta}{\sqrt{2}}\,.
\end{equation}

Our azimuthal angles $\varphi_{1(2)}$ are similar to those defined in Higgs to $4$ leptons decay analyses~\cite{Sirunyan:2017tqd,ATLAS:2017cju}. There is however one important difference, namely the fact that their orientations have been specified in terms of fermions of given helicities, while fermions are distinguished by their electric charge in the standard definition. We are obliged to work in this non-standard formalism by the fact that the orientation has to match the one employed in the partial wave decomposition formula, where the ordering of the fermions determines whether $\lambda=\lambda_1-\lambda_2$ equals plus or minus one. Concretely this makes a difference only if we consider the leptonic decay of the $Z$ boson, where the charge of the final state fermions is measurable while their helicity of course is not. Therefore while the standard  angles (defined with fermion charge orientation) can be fully determined in this case, our angles are subject to the discrete ambiguity $\{\theta_{1(2)},\varphi_{1(2)}\}\leftrightarrow \{\pi-\theta_{1(2)},\varphi_{1(2)}+\pi\}$ resulting from the inability to tell right-handed from left-handed leptons. Notice that the left-right ambiguity we just described does not arise in the case of a leptonically decaying $W$ boson, where fermion chirality is known theoretically in terms of the electric charge of the charged lepton. 
However in this case determining the decay angles requires the reconstruction of the neutrino momentum, 
 that introduces a reconstruction ambiguity corresponding approximately (for boosted $W$) to $\{\theta_{1(2)},\varphi_{1(2)}\}\leftrightarrow \{\theta_{1(2)},\pi-\varphi_{1(2)}\}$. We will discuss this in detail in section~\ref{sec:Wreco}.
 
Taking the modulus square of eq.~(\ref{amp}) we obtain interference terms between diboson amplitudes of different helicities. Denoting as ${\mathbf{h}}=(h_1,h_2)$ the  vector formed by the two boson helicities and by ${\mathbf{\Delta h}}={\mathbf{h^\prime}}-{\mathbf{h}}$ the helicity difference between the two interfering amplitudes, the interference terms can be synthetically written as 
\begin{equation}\label{int0}
I_{{\mathbf{h}}\otimes {\mathbf{h^\prime}}}^{V_1V_2}=T_{{\mathbf{h}}{\mathbf{h^\prime}}}^{V_1V_2}|{\mathcal{A}}_{{\mathbf{h}}}{\mathcal{A}}_{{\mathbf{h^\prime}}}|\cos{[{\mathbf{\Delta h}}\cdot{\bm{\varphi}}+\delta]}\,,
\end{equation}
where ${\bm{\varphi}}=(\varphi_1,\varphi_2)$ and $\delta$ is the relative phase between ${\mathcal{A}}_{{\mathbf{h}}}$ and ${\mathcal{A}}_{{\mathbf{h^\prime}}}$, i.e. ${\mathcal{A}}_{{\mathbf{h}}}{\mathcal{A}}_{{\mathbf{h^\prime}}}^*=|{\mathcal{A}}_{{\mathbf{h}}}{\mathcal{A}}_{{\mathbf{h^\prime}}}| e^{i\delta}$. The dependence on $\theta_{1(2)}$ and the couplings has been encapsulated in 
\begin{equation}
T_{{\mathbf{h}}{\mathbf{h^\prime}}}^{V_1V_2}=2\,g_1^2g_2^2 d_{h_1}d_{h_1^\prime}d_{h_2}d_{h_2^\prime}\,.
\end{equation}
The equation reproduces the well-known result according to which intermediate particles of different helicities do interfere, a priori, and that integration over the two azimuthal decay angles is needed in order to cancel the interference term and obtain a factorised production-times-branching-ratio cross-section. Notice that instead integrating over the polar angles $\theta_{1,2}$ does not cancel the interference because the Wigner $d$-functions are not orthogonal. This means that while the interference effects we are interested in are present in the data, it is not hard to ``kill'' them by measuring only quantities that are inclusive on the azimuthal angles. Examples of such quantities are the reconstructed momenta of the two bosons and the variables (transverse or invariant mass, $p_T$ or rapidity) obtained out of them, which are normally employed in experimental analyses of diboson processes. Interference resurrection requires measuring the azimuthal angles $\varphi_{1(2)}$, or other kinematical variables that are sensitive to those.

Before moving to concrete examples of interference resurrection, it is worth noticing that eq.~(\ref{int0}) can be further simplified if we restrict ourselves to $2\rightarrow2$ diboson processes at tree-level, by exploiting an interesting connection with  $CP$ symmetry. The point is that the complex conjugate of a tree-level amplitude that receives no contribution from nearly on-shell intermediate resonance exchange is equal to the amplitudes for the ``reversed'' process with $in$ and $out$ states interchanged. This result is a consequence of the Optical Theorem and applies to SM amplitudes as well as amplitudes induced by EFT operators. The reversed amplitude is in turn related to the original amplitude by time-reversal symmetry, which acts on the amplitudes like $CP$ because of the $CPT$ theorem. With these elements one can prove\footnote{The simplest way is to employ the partial-wave decomposition of the amplitude, noticing that time-reversal acting on partial wave amplitudes just exchanges $in$ and $out$ states without extra phases~\cite{Jacob:1959at}. Using the optical theorem, and the fact that the basis functions in the amplitude decomposition become real if the scattering occurs (as it does in the special frame) on the $x$-$z$ plane, the result is immediately derived.} that the $2\rightarrow2$ amplitudes evaluated in the special frame obey
\begin{equation}\label{CP}
\left({\mathcal{A}}_{{\mathbf{h}}}\right)^*=\rho_{CP}{\mathcal{A}}_{{\mathbf{h}}}\,,
\end{equation}
where $\rho_{CP}=+ 1$ or $-1$ for $CP$-preserving and $CP$-violating amplitudes, respectively. We are interested in the interference between the SM term, which is $CP$-even, and BSM contributions to the amplitude originating from EFT operators that are either $CP$-even or $CP$-odd. In the former case the product of the two interfering amplitudes is purely real, while it is purely imaginary in the latter one. Eq.~(\ref{int0})  thus becomes
\begin{eqnarray}\label{intmm}
\hspace{-15pt}&&\hspace{-4pt}I_{{\mathbf{h}}\otimes {\mathbf{h^\prime}}}^{V_1V_2}\hspace{-2pt}=\hspace{-2pt}
T_{{\mathbf{h}}{\mathbf{h^\prime}}}^{V_1V_2}\hspace{-3pt}
\left[{\mathcal{A}}_{{\mathbf{h}}}^{\textrm{\tiny{SM}}}{\mathcal{A}}_{{\mathbf{h^\prime}}}^{\textrm{\tiny{BSM}}_+}\hspace{-4pt}
+{\mathcal{A}}_{{\mathbf{h}}}^{\textrm{\tiny{BSM}}_+}{\mathcal{A}}_{{\mathbf{h^\prime}}}^{\textrm{\tiny{SM}}}
\right]\hspace{-2pt}
\cos{[{\mathbf{\Delta h}}\cdot{\bm{\varphi}}]}\,,\\
\hspace{-15pt}&&\hspace{-4pt}I_{{\mathbf{h}}\otimes {\mathbf{h^\prime}}}^{V_1V_2}\hspace{-2pt}=\hspace{-2pt}
iT_{{\mathbf{h}}{\mathbf{h^\prime}}}^{V_1V_2}\hspace{-3pt}
\left[{\mathcal{A}}_{{\mathbf{h}}}^{\textrm{\tiny{SM}}}{\mathcal{A}}_{{\mathbf{h^\prime}}}^{\textrm{\tiny{BSM}}_-}\hspace{-4pt}
-{\mathcal{A}}_{{\mathbf{h}}}^{\textrm{\tiny{BSM}}_-}{\mathcal{A}}_{{\mathbf{h^\prime}}}^{\textrm{\tiny{SM}}}
\right]\hspace{-2pt}
\sin{[{\mathbf{\Delta h}}\cdot{\bm{\varphi}}]}\,,\nonumber
\end{eqnarray}
for $CP$-even and $CP$-odd BSM physics, respectively. Interestingly enough, measuring if the interference assumes sine or cosine form (or a combination of the two, if both effects are present) would allow us to distinguish CP-even from CP-odd new physics effects. 

So far we have discussed massive diboson production; if one of the bosons is a photon, the result is simpler and can be worked out along similar lines. In this case we have only one polar and one azimuthal angle $\theta$ and $\varphi$ associated with the decay of the massive boson $V$. The photon is a real final-state particle of helicity $h_\gamma=\pm1$ and no interference is possible between different photon helicity amplitudes. Furthermore, at high-energy, the only relevant BSM effects emerge in amplitudes where $V$ is also transverse, since amplitudes with only one longitudinal vector boson are suppressed by $m_V/E$. The interference among $V$-helicity configurations $h=\pm1$ and $h^\prime=\mp1$ reads in this case
\begin{eqnarray}\label{intmg}
\hspace{-15pt}&&\hspace{-10pt}I_{{{h}}\otimes {{h^\prime}}}^{V\gamma,h_\gamma}\hspace{-5pt}=\hspace{-2pt}2g^2\hspace{-2pt}\sin^2\hspace{-1pt}\theta \hspace{-3pt}
\left[\hspace{-1pt}{\mathcal{A}}_{h\,h_\gamma}^{\textrm{\tiny{SM}}}\hspace{-2pt}{\mathcal{A}}_{h^\prime\,h_\gamma}^{\textrm{\tiny{BSM}}_+}
\hspace{-4pt}+\hspace{-2pt}{\mathcal{A}}_{h\,h_\gamma}^{\textrm{\tiny{BSM}}_+}\hspace{-2pt}{\mathcal{A}}_{h^\prime\,h_\gamma}^{\textrm{\tiny{SM}}}\hspace{-2pt}
\right]\hspace{-2pt}\cos{\Delta h\varphi}\,,\\
\hspace{-15pt}&&\hspace{-10pt}I_{{{h}}\otimes {{h^\prime}}}^{V\gamma,h_\gamma}\hspace{-5pt}=\hspace{-2pt}2ig^2\hspace{-2pt}\sin^2\hspace{-1pt}\theta \hspace{-3pt}
\left[\hspace{-1pt}{\mathcal{A}}_{h\,h_\gamma}^{\textrm{\tiny{SM}}}\hspace{-2pt}{\mathcal{A}}_{h^\prime\,h_\gamma}^{\textrm{\tiny{BSM}}_-}
\hspace{-4pt}-\hspace{-2pt}{\mathcal{A}}_{h\,h_\gamma}^{\textrm{\tiny{BSM}}_-}\hspace{-2pt}{\mathcal{A}}_{h^\prime\,h_\gamma}^{\textrm{\tiny{SM}}}\hspace{-2pt}
\right]\hspace{-2pt}\sin{\Delta h\varphi}\,,\nonumber
\end{eqnarray}
where $\Delta{h}=h^\prime-h$.

In the rest of the paper we work out concrete example of such interference effects, focussing with greater details on the $W\gamma$ production process with leptonically decaying~$W$. The most difficult part of the analysis will be the reconstruction of the $W$ boson decay angle, a technique that furthermore can be useful for interference resurrection also in other channels. We thus discuss it extensively in the next section.

\section{Leptonic $W$ Reconstruction}\label{sec:Wreco}
Measuring $\varphi$ is essential for interference resurrection, as explained above. This would be relatively straightforward for hadronically decaying $W$ boson \footnote{We implicitly assume here that momenta of all the other particles produced in association with the $W$ can be measured directly.}, but the difficulties of boosted hadronic $W$ tagging and QCD background suppression make the leptonic case simpler to study (see however section~\ref{sec:otherchannels}). In the leptonic case, determining the $W$ decay angles $\theta$ and $\varphi$  requires instead neutrino momentum reconstruction. This is performed with the standard strategy of identifying the neutrino transverse momentum (${\vec{p}}_{\bot \nu}$) with the missing transverse energy vector (${\vec{E}}_{\bot}^{\rm{miss}}$) and determining the neutrino rapidity by imposing that the lepton-neutrino invariant mass equals the $W$ pole mass $m_W$. If the lepton and the neutrino emerge from a virtual $W$ with mass exactly equal to $m_W$, the equation has two solutions
\begin{eqnarray}\label{numon}
\hspace{-25pt}&&\eta_\nu^{\pm}\hspace{-2pt}-\hspace{-2pt}\eta_l\hspace{-2pt}=\hspace{-2pt}\pm \left|\cosh^{-1}[1+\Delta^2]\right|\hspace{-2pt}=\hspace{-2pt}\pm\log\hspace{-3pt}\left[1\hspace{-3pt}+\hspace{-2pt}\Delta\sqrt{2\hspace{-2pt}+\hspace{-2pt}\Delta^2}+\hspace{-2pt}\Delta^2\hspace{-1pt}\right]\,,\nonumber\\
\hspace{-25pt}&&\Delta^2=\frac{m_W^2-m_{\bot}^2}{2 p_{\bot l}p_{\bot \nu}}\,,
\end{eqnarray}
where $m_{\bot}<m_W$ is the $W$-boson transverse mass. Only one of the two solutions reproduces the true neutrino momentum, and there is no way to tell which one. We thus decided to pick one of these solutions at random on an event-by-event basis. This introduces, even before detector effects are taken into account, an uncertainty in the determination of the neutrino momentum.

We focus in particular on the boosted $W$ regime $\Delta\ll 1$, which is the relevant one for our high-energy analysis. Neutrino reconstruction becomes, from a purely theoretical viewpoint, increasingly accurate in this limit because the two solutions for $\eta_\nu$ in \eq{numon} tend to coincide, $\eta_\nu^{\pm}=\eta_l\pm\sqrt{2}\Delta+O(\Delta^3)\to \eta_l$. However an interesting subtlety emerges, related with the fact that not all the quantities computed on the two neutrino momentum solutions coincide in the limit, but only the four components of the reconstructed $W$ momentum and the polar decay angle $\theta$. For the $W$ momentum, which is just the sum of the lepton and of the reconstructed neutrino momenta, this is rather obviously the case. The fact that the solutions give coincident $\theta$ can be seen by recalling the standard kinematics of nearly massless parton splitting, that gives $1+\cos\theta\simeq2x$, where $x=E_l/E_W\simeq p_{\bot l}/(p_{\bot l}+p_{\bot \nu})$ is the $W$ energy fraction carried away by the charged lepton in the splitting. Since $\theta$, in the limit, becomes function of observed components ($p_{\bot l}$ and $p_{\bot \nu}$) only, it must be the same when evaluated on the two solutions. The situation is instead very different for $\varphi$, that does not converge to a unique value.  In the large-boost expansion $m_W^2/p_{\bot l}p_{\bot \nu}\ll1$ we find \footnote{This implicitly assumes a $2\rightarrow2$ process.}
\begin{equation}\label{phireq}
\hspace{-4pt}\cot\hspace{-1pt}\varphi\hspace{-1pt}=\hspace{-2pt}\frac1{\sin\hspace{-2pt}{[\phi_\nu\hspace{-2pt}-\hspace{-2pt}\phi_l]}}\left[\sinh[\eta_l\hspace{-2pt}-\hspace{-2pt}\eta_\nu]\hspace{-2pt}+\hspace{-2pt} {\mathcal{O}}\hspace{-2pt}\left(\hspace{-1pt}\frac{m_W^2}{p_{\bot l}p_{\bot \nu}}\hspace{-1pt}\right)\right]\,,
\end{equation}
where $\phi_{l}$ and $\phi_\nu$ are the lepton and neutrino azimuthal coordinates in the lab frame. When evaluated on the two solutions $\eta_\nu=\eta_\nu^\pm$ (see eq.~(\ref{numon})), the first term in the square bracket tends to zero as $\Delta\sim m_W/p_{\bot}$ in the boosted limit, and dominates over the second one. This term has opposite sign on the two solutions. In the boosted limit, the lepton and the neutrino become close to each other also in the transverse plane, therefore $\phi_\nu-\phi_l$ tends to zero and it is possible to show that it scales like $\Delta$. Eq.~(\ref{phireq}) thus goes to a constant in the limit, producing two opposite values for $\cot\varphi^\pm$ when computed in the two solutions. 
So, only one of the two reconstructed values of $\varphi$ will be close to the true decay angle, the other one will be ${\mathcal{O}}(1)$ different, and related to the former by a discrete operation under which the cotangent changes sign. Since it is possible to show that the sine of $\varphi$ approaches a unique limit, the relation among $\varphi^+$ and $\varphi^-$ is 
\begin{equation}\label{ambiguityW}
\varphi^+=\pi-\varphi^- \mod 2\pi\,,
\end{equation}
and produces the reconstruction ambiguity mentioned in section~2.
\begin{figure*}[ht!]
\centering
\includegraphics[width=0.45\textwidth]{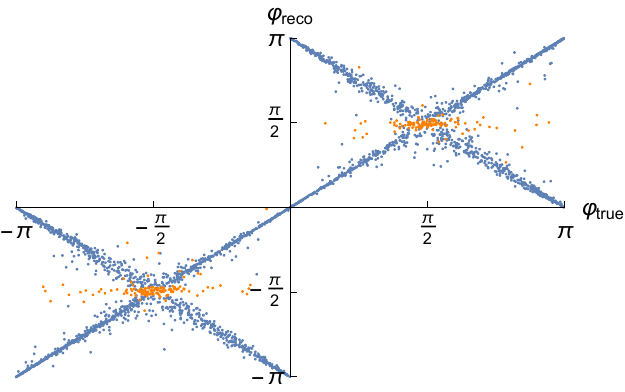}\hspace{1cm}
\includegraphics[width=0.45\textwidth]{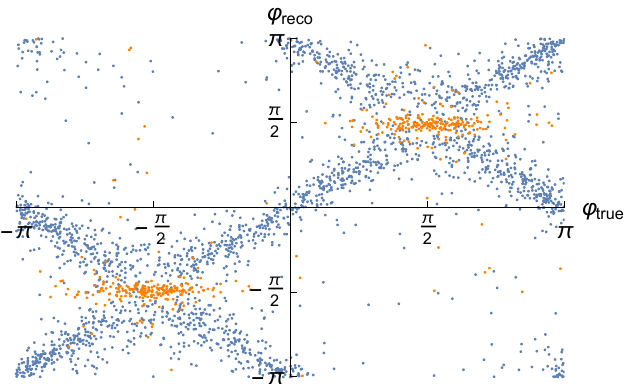}
\caption{\it Correlation between the real  and  reconstructed angles $\varphi_{true}$, $\varphi_{reco}$ in $W^+\gamma$ $(W^+\to e^+\nu)$ processes, without detector effects (left) and including detector effects, simulated with {\sc{Delphes}} (right).
Events are selected if they pass the following selection criteria: $p_{\bot \gamma}>300\ {\rm{GeV}}$, $p_{\bot l},{{E}}_{\bot}^{\rm{miss}}>80$~GeV, $\Delta R(\gamma,l)>3$ and $\eta_l<2.4$. Points with a unique solution $\varphi^+=\varphi^-$ are highlighted in orange.}
\label{delphes}
\end{figure*}

In reality, the virtual $W$ mass is not exactly $m_W$, but is very close to that because the $W$ is narrow. The ``right'' solution will thus provide a good approximation of the true kinematics. However, experimental errors in the measurement of the lepton momentum or of ${\vec{E}}_{\bot}^{\rm{miss}}$, or the fact that the virtual $W$ mass was truly slightly above $m_W$, can lead to events with $m_{\bot}>m_W$. Then \eq{numon} has no real solution and the neutrino is reconstructed by requiring that the lepton-neutrino invariant mass is as close as possible to~$m_W$. This selects a unique configuration $\eta_\nu=\eta_l$. In this situation, the first term in the square bracket of eq.~(\ref{phireq}) is exactly zero, $\phi_\nu-\phi_l$ scales like $\Delta$ in the boosted limit as previously mentioned, while the second term in the square bracket vanishes as $\Delta^2$. The reconstructed $\varphi$ thus approaches a configuration with $\cot\varphi=0$, corresponding to 
\begin{equation}\label{badsol}
 \varphi=\pi/2 \quad\textrm{or}\quad\varphi=-\pi/2. 
\end{equation}
The other variables, namely the $W$ momentum and $\theta$, are instead correctly reproduced in the limit.

The peculiar behaviour of the reconstructed $\varphi$, summarized in eqs.~(\ref{ambiguityW},\ref{badsol}), is illustrated  in fig.~\ref{delphes}, where the true $\varphi_{true}$ is compared with the reconstructed one $\varphi_{\rm{reco}}$ in the example of $W\gamma$ final states (that is relevant for the analysis of the next section). We have selected events with photon transverse momenta $p_{\bot\gamma}>300$~GeV, while $p_{\bot l},{{E}}_{\bot}^{\rm{miss}}>80$~GeV in order to avoid pathological cases where one of the final state leptons is extremely soft. Generation-level ({\sc MadGraph}~\cite{Alwall:2014hca},~\cite{Degrande:2012wf}) events are shown on the left panel while {\sc{Delphes}}~\cite{deFavereau:2013fsa} detector effects are included in the right one (Pythia~8~\cite{Sjostrand:2014zea} is used for showering and hadronization). If $m_{\bot l}<m_W$ (blue points) we take one of the two solutions at random as previously discussed, however we verified that the figure (and the rest of the analysis) would not change if we had taken systematically the $+$ or the $-$ solution. The events where $m_{\bot l}>m_W$, marked in orange, mostly give a reconstructed angle of $\pm\pi/2$, often also in events where the true angle was far from $\pm\pi/2$. Detector resolution has a considerable impact on the determination of $\varphi$, as it was to be expected because in the boosted regime the lepton and the neutrino get close to each other and the determination of the scattering plane becomes increasingly sensitive to uncertainties in ${\vec{E}}_{\bot}^{\rm{miss}}$ and in the lepton momentum. Notice also that detector effects populate the $m_{\bot l}>m_W$ region, making indeed more orange points appear in the figure. This induces an anomalous concentration of points at $\varphi_{\rm{reco}}\sim\pm\pi/2$.

\section{Anomalous Gauge Couplings in $W\gamma$}

The only $d=6$ EFT  operators that give unsuppressed high-energy contributions to the ${{W\gamma}}$ channel are (with the conventions of ref.~\cite{Grzadkowski:2010es})
\begin{equation}
{\cal O}_{3W}=\epsilon^{ijk}W_\mu^{i \nu}W_\nu^{j\rho}W_\rho^{k \mu}\,,\quad
{\cal O}_{3\widetilde W}=\epsilon^{ijk}\widetilde W_\mu^{i \nu}W_\nu^{j\rho}W_\rho^{k \mu}\,,\nonumber
\end{equation}
that are respectively $CP$-even and $CP$-odd, and correspond to modifications of the trilinear gauge couplings of
ref.~\cite{Hagiwara:1986vm}, as $\lambda_\gamma=6 C_{3W} m_W^2/g$ (and similarly for CP-odd quantities), where ${C}_i$ are the coefficients, with energy dimension $-2$, appearing in the Lagrangian as ${\cal L}_{\tiny \textrm{BSM}}=\sum{C}_{i} {{\cal O}}_{i}$. At high energy they give a quadratically enhanced contribution only to same-helicity ${{W\gamma}}$ final states, namely
\begin{eqnarray}
&&{\mathcal{A}}_{+\,+}^{\textrm{\tiny{BSM}}_+}={\mathcal{A}}_{-\,-}^{\textrm{\tiny{BSM}}_+}\approx C_{3W} 6e\sqrt{2}M^2_{W\gamma} \sin\Theta \,,\nonumber\\
&&{\mathcal{A}}_{+\,+}^{\textrm{\tiny{BSM}}_-}=-{\mathcal{A}}_{-\,-}^{\textrm{\tiny{BSM}}_-}\approx  i C_{3\widetilde W} 2e \sqrt{2}M^2_{W\gamma} \sin\Theta \,,
\end{eqnarray}
where $\Theta$ is the diboson scattering angle and $M_{W\gamma}$ the invariant mass of the $W\gamma$ system; $e$ is the electric charge. Their contribution is instead not enhanced in the opposite-helicity channel, which on the other hand is the only sizable one in the SM, where ${\mathcal{A}}_{\pm\,\pm}^{\textrm{\tiny{SM}}}\sim m_W^2/M_{W\gamma}^2$. This fact is the essence of the non-interference problem~\cite{Azatov:2016sqh} mentioned in the introduction. By eq.~(\ref{intmg}), after summing over the photon polarizations (which are not observable), we obtain 
\begin{eqnarray}\label{intWg}
&&I_{{{-}}\otimes {{+}}}^{W\gamma}\hspace{-2pt}=\hspace{-2pt}2g^2\hspace{-2pt}\sin^2\hspace{-1pt}\theta
{\mathcal{A}}_{++}^{\textrm{\tiny{BSM}}_+}\hspace{-2pt}
\left[\hspace{-1pt}{\mathcal{A}}_{-+}^{\textrm{\tiny{SM}}}\hspace{-4pt}+\hspace{-2pt}{\mathcal{A}}_{+-}^{\textrm{\tiny{SM}}}\hspace{-2pt}
\right]\hspace{-2pt}\cos{2\varphi}\,,\nonumber\\
&&I_{{{-}}\otimes {{+}}}^{W\gamma}\hspace{-2pt}=\hspace{-2pt}2ig^2\hspace{-2pt}\sin^2\hspace{-1pt}\theta
{\mathcal{A}}_{++}^{\textrm{\tiny{BSM}}_-}\hspace{-2pt}
\left[\hspace{-1pt}{\mathcal{A}}_{-+}^{\textrm{\tiny{SM}}}\hspace{-4pt}-\hspace{-2pt}{\mathcal{A}}_{+-}^{\textrm{\tiny{SM}}}\hspace{-2pt}
\right]\hspace{-2pt}\sin{2\varphi}\,.
\end{eqnarray}
By looking at these equations one might worry about possible cancellations, occurring in one of the two interference terms, in the presence of exact or approximate relations between the $(-+)$ and $(+-)$ SM amplitudes. However no such relations exist and the two interference terms are of comparable magnitude once integrated over the diboson scattering angle $d\cos\Theta$. 
\begin{figure*}
\centering
\includegraphics[width=0.45\textwidth]{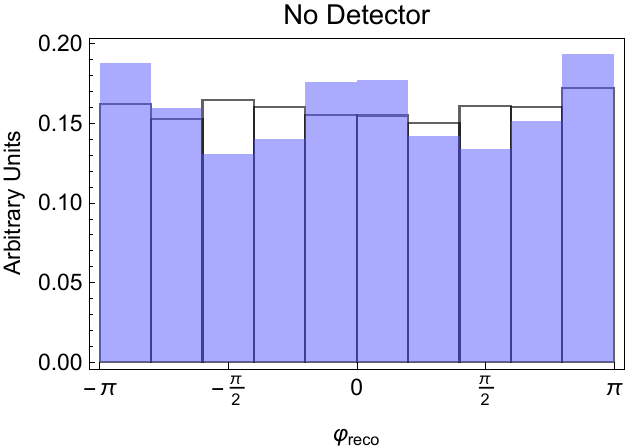}\hspace{1cm}
\includegraphics[width=0.45\textwidth]{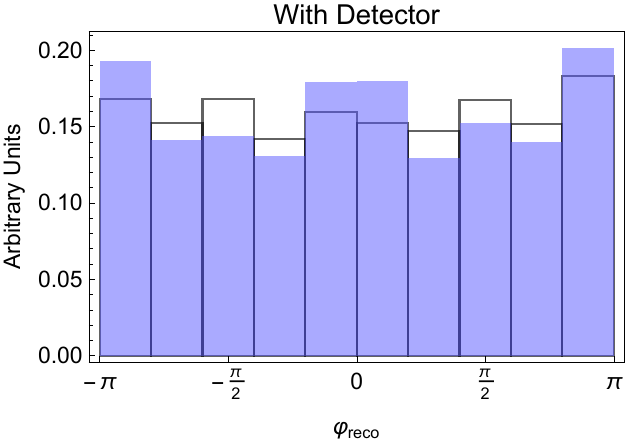}
\caption{\it Reconstructed azimuthal angular distribution in the SM (black lines) and BSM (blue area, with $C_{3W}=0.2$ TeV$^{-2}$), normalized to unity with {\sc Delphes} detector study (right) and without (left). Same selection cuts as fig.~\ref{delphes}.}
\label{fig:dist}
\end{figure*}

Following our discussion in section~\ref{sec:Wreco}, we should  average our interference formula (\ref{intWg}) over the two ambiguous configurations in \eq{ambiguityW}, obtaining the following interesting result. Interference with $CP$-odd new physics ${\cal O}_{3\widetilde W}$, is opposite in the two ambiguous configurations, therefore it cancels in the average giving us no chance to detect it in the $W\gamma$ final state. Interference with $CP$-even new physics ${\cal O}_{3W}$, is instead invariant under $\varphi\rightarrow\pi-\varphi$, hence it is unaffected by the average and perfectly visible in spite of the ambiguity. This is verified in fig.~\ref{fig:dist}, where we show the reconstructed $\varphi$ distribution with the same cuts of fig.~\ref{delphes}. The SM is nearly flat, as expected\footnote{In fact, even in the SM, interference between the $\pm\mp$ and the longitudinal $0\mp$ amplitudes -- which are suppressed by only one power of the energy in the boosted regime -- induces a mild $\sim \cos\varphi$ behaviour, that is however invisible due to the reconstruction ambiguity of \eq{ambiguityW}.}, while BSM (taking $C_{3W}=0.2$ TeV$^{-2}$ for illustration) introduces a $\cos2\varphi$ behaviour. The little bumps at $\pm\pi/2$ are due to $m_{\bot}>m_W$ configurations. Aside from those, the effect of the {\sc{Delphes}} smearing on the distribution is mild.

The rest of the analysis is straightforward. We simulate leptonic decays of $W^+\gamma$ where, in addition to the cuts for fig.~\ref{delphes}, we consider $p_{\bot\gamma}$ bins of \{150, 210, 300, 420, 600, 850, 1200\} GeV, increasing linearly in size to accomodate experimental resolution on $p_{\bot\gamma}$, but as fine as possible to maximize the sensitivity to BSM effects.  In addition, we consider 10 azimuthal angular bins $\in[-\pi,\pi]$, where we fit the number of events to a quadratic function of $C_{3W}$. We repeat the simulation with and without {\sc Delphes} detector simulation, to quantify the impact of these effects. Notice that when quoting generator-level results, we take into account an overall reconstruction efficiency $\sim 0.6$ extracted from the comparison with {\sc Delphes}. Reducible backgrounds are not taken into account in the simulation, in spite of the fact that jets faking photons give nearly $50\%$ of the SM $W\gamma$ contribution in existing run-$1$ studies of the $W\gamma$ final state \cite{Chatrchyan:2013fya}. However ref.~\cite{Chatrchyan:2013fya} focuses on lower photon momenta ($p_{\bot\gamma}\lesssim200$GeV) than those that are relevant for our analysis. We thus expect the jet background to be less relevant in our case because the photon mistag rate for jets decreases with $p_{\bot\gamma}$ \cite{ATLAS_Photon} and because the $Wj$ cross section should decrease faster than $W\gamma$ due to the steeply falling gluon parton distribution function. Still, we expect this background to be significant.

\begin{figure*}
\centering
\includegraphics[width=0.45\textwidth]{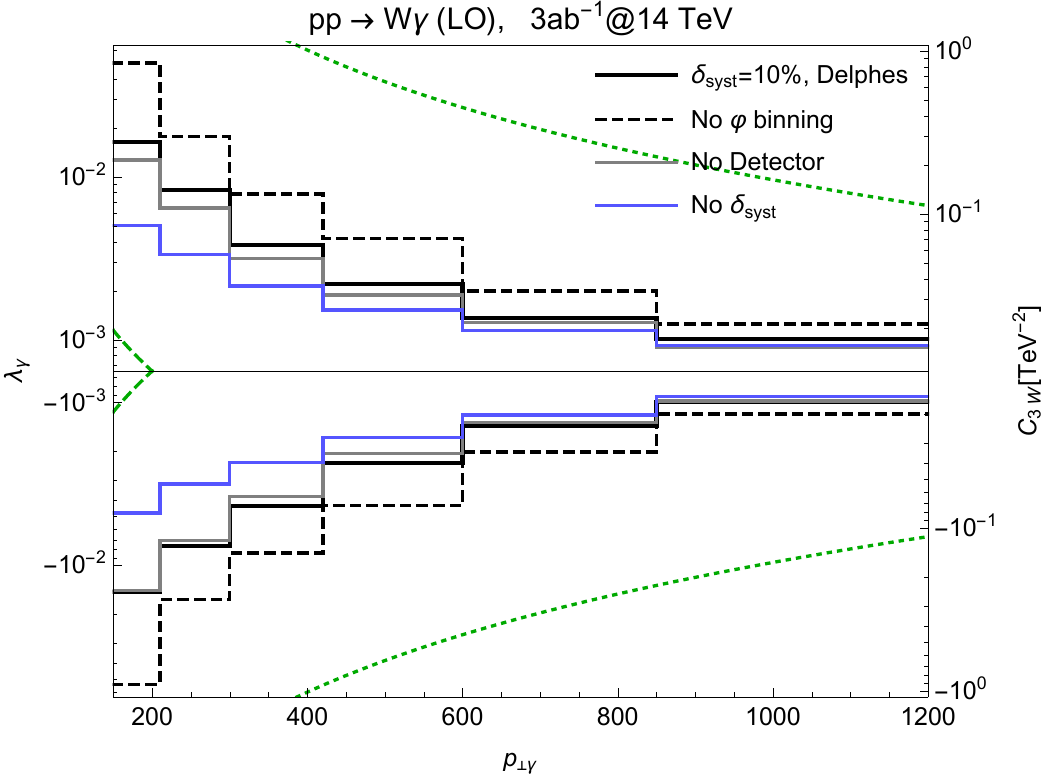}\hspace{1cm}
\includegraphics[width=0.45\textwidth]{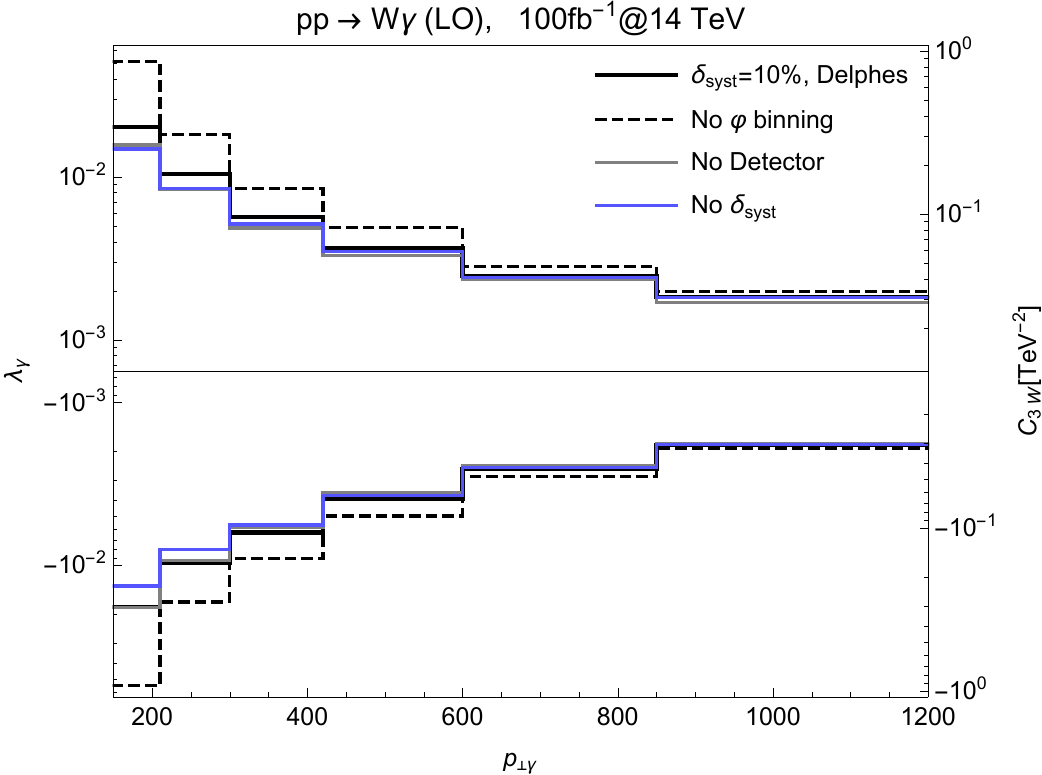}
\caption{\it Projected 95\%C.L. sensitivity on $\lambda_\gamma$ and $C_{3W}$ for the 14 TeV LHC with an integrated luminosity of 3ab$^{-1}$ (left) or 100fb$^{-1}$ (right), in different cases. Green dotted and dashed curves correspond to $C_{3W}=g/M^2$ and $C_{3W}=g^3/(16\pi^2M^2)$ ($M\approx 2p_{\bot\gamma}$) respectively. }
\label{fig:fit}
\end{figure*}

The results are shown in fig.~\ref{fig:fit}, in terms of the projected sensitivity at the end of the High-Luminosity LHC program (3ab$^{-1}$, left panel) and at an earlier stage (100fb$^{-1}$, right panel). The left vertical axis shows the reach in terms of anomalous couplings $\lambda_\gamma$ while the right axis is expressed in terms of $C_{3W}$. As in ref.~\cite{Farina:2016rws}, we show how the reach deteriorates when high-energy (high-$p_{\bot\gamma}$) bins are ignored in the fit, with the aim of outlining which kinematical regime ($p_{\bot\gamma}\lesssim1$~TeV, in this case) is relevant for the limit. Accurate experimental measurements are needed in this regime, together with a trustable EFT prediction, i.e. an EFT cutoff $\Lambda>1$~TeV.\footnote{This way of assessing the EFT validity was advocated in \cite{Biekoetter:2014jwa,Busoni:2014sya,Racco:2015dxa,Contino:2016jqw}.} The full simulation, with a 10\% systematic relative uncertainty, summed in quadrature with the statistical one, is portrayed in black in the figure, while the analogous analysis, but without binning in the azimuthal angle $\varphi$, is shown dashed. The comparison of these two lines shows the added value of our analysis. Detector effects can be quantified instead by comparing with the gray line, while the impact of systematic errors is captured by comparison with the blue line. 

For reference, we also show in green (dotted, dashed) theoretical curves corresponding to different power countings, $C_{3W}=g/\Lambda^2$ and $C_{3W}=g^3/(16\pi^2\Lambda^2)$, reflecting different BSM hypotheses, see ref.~\cite{Liu:2016idz}. Here we approximate $\Lambda\simeq 2 p_{\bot\gamma}$ to argue that, for models that reflect the first power counting (dotted curve), the bounds we obtain are well within the EFT validity, in all transverse-momentum bins. For weakly coupled models, where these effects arise at loop-level (dashed curve), the projected sensitivity is instead not enough.
A popular heuristic method to assess the validity of the EFT expansion is to present results with and without the BSM-squared contributions in the cross-section. We have checked that, with this procedure, bounds without interference resurrection deteriorate by one order of magnitude, while interference-resurrection bounds are much more stable.

Our analysis could be improved by considering additional variables, such as the polar angle $\theta$. A central cut in $\theta$ would indeed enhance the interference term (\ref{intmg}) compared to the non-interference ones that are proportional to $d_\pm^2$ and are thus preferentially forward or backward. We could also exploit the dependence on $\Theta$, which we could readily get from eq.~(\ref{intWg}). We leave this for future work.

\section{Other Channels}\label{sec:otherchannels}

Interference resurrection could be useful in all channels where BSM effects hide in vector boson final states that are rare in the SM. In this section we mention some other interesting applications.

\vspace{.5em}
\noindent {\bf Hadronic ${\mathbf{W}}$ decays and CP-odd effects.} $CP$-odd new physics cannot be detected in the leptonic $W\gamma$ channel because of the reconstruction ambiguity. In hadronic channels the ambiguity in the determination of $\varphi$ comes instead from the inability to measure  final state quarks flavour. Considering for instance a charge-plus $W$, we cannot tell which one of the jets (or subjets, since the $W$ is boosted) is the down anti-quark (which has necessarily $+1/2$ helicity) and which one is the up quark. If one of the two is picked at random to be the $\bar{d}$, we should average the cross-section over the left-right ambiguity $\{\theta_{1(2)},\varphi_{1(2)}\}\leftrightarrow \{\pi-\theta_{1(2)},\varphi_{1(2)}+\pi\}$ described in section~\ref{sec:IR} in the context of $Z$-decays. 
These operations leaves both $CP$-even and $CP$-odd interferences (\ref{intmg}) invariant, hence it does not prevent their observability. Clearly in the hadronic case we do not even know the $W$ charge, however summing over charges cannot produce a cancellation because the total cross-sections for $W^+$ and $W^-$ are different at order one.
This remains a challenging channel, because of the need of boosted hadronic $W$ reconstruction and because of the important QCD backgrounds.

\vspace{.5em}
\noindent{\bf Fully leptonic ${\mathbf{WZ}}$.} Despite the smaller rate, due to the small leptonic $Z$ branching ratio, here we can integrate over the $W$ decay angles, which we studied already in leptonic $W\gamma$, and focus on the $Z$ decay angles $\theta_\textrm{\tiny Z}$~and~$\varphi_\textrm{\tiny Z}$. As discussed 
in section~\ref{sec:IR} it is convenient to think in terms of the ``standard''  angles $\theta_\textrm{\tiny Z}^c$ and $\varphi_\textrm{\tiny Z}^c$, with orientations defined in term of the charge-plus final state lepton rather than of the one with $+1/2$ helicity.
Standard angles are fully measurable, but they are related with ``our angles'' $\theta_\textrm{\tiny Z}$ and $\varphi_\textrm{\tiny Z}$ in a way that depends on the chirality of the $Z$ boson decay. Namely, if the $Z$ decays to left-handed spinors, the standard angles coincide with ours, otherwise they are related to the former by the left-right ambiguity (section~\ref{sec:IR}) $\{\theta_{1(2)},\varphi_{1(2)}\}\leftrightarrow \{\pi-\theta_{1(2)},\varphi_{1(2)}+\pi\}$. When computing the differential cross-section in the $\theta_\textrm{\tiny Z}^c$ and $\varphi_\textrm{\tiny Z}^c$ angles we should thus sum over the two ambiguous configurations, taking of course into account that the left-handed $Z$ decay coupling, $g_L$, is different from the right-handed one $g_R$. The result is readily obtained from eq.~(\ref{intmm}) and reads
\begin{eqnarray}\label{intWZ}
\hspace{-20pt}&&I_{{{-}}\otimes {{+}}}^{WZ}\hspace{-2pt}=\hspace{-2pt}2[g_L^2+g_R^2]\sin^2\hspace{-1pt}\theta_\textrm{\tiny Z}^c
{\mathcal{A}}_{++}^{\textrm{\tiny{BSM}}_+}\hspace{-2pt}
\left[\hspace{-1pt}{\mathcal{A}}_{-+}^{\textrm{\tiny{SM}}}\hspace{-4pt}+\hspace{-2pt}{\mathcal{A}}_{+-}^{\textrm{\tiny{SM}}}\hspace{-2pt}
\right]\hspace{-2pt}\cos{2\varphi_\textrm{\tiny Z}^c}\,,\nonumber\\
\hspace{-20pt}&&I_{{{-}}\otimes {{+}}}^{WZ}\hspace{-2pt}=\hspace{-2pt}2i[g_L^2+g_R^2]\sin^2\hspace{-1pt}\theta_\textrm{\tiny Z}^c
{\mathcal{A}}_{++}^{\textrm{\tiny{BSM}}_-}\hspace{-2pt}
\left[\hspace{-1pt}{\mathcal{A}}_{-+}^{\textrm{\tiny{SM}}}\hspace{-4pt}-\hspace{-2pt}{\mathcal{A}}_{+-}^{\textrm{\tiny{SM}}}\hspace{-2pt}
\right]\hspace{-2pt}\sin{2\varphi_\textrm{\tiny Z}^c}\,.\hspace{2.5em}
\end{eqnarray}
Since it is invariant under the left-right ambiguity, $h_\textrm{\tiny Z}=+1$ interference with $h_\textrm{\tiny Z}=-1$ does not cancel in the sum, neither in the $CP$-even nor in the $CP$-odd BSM case. This in principle would allow us to detect $CP$-odd interference. Notice that in the $WZ$ channel one could also take study the interference of BSM effects with the longitudinal-longitudinal SM amplitude ${\mathcal{A}}_{00}^{\textrm{\tiny{SM}}}$, which is non-vanishing in the high-energy limit. These effects cancel if we integrate over the $W$ boson azimuthal angle, and do not carry radically new information on new physics. Hence can be safely ignored, at least at a first stage.

\vspace{.5em}
\noindent{\bf New physics  in the longitudinal polarizations.} As a matter of fact, longitudinal polarizations, though surviving in the high-energy limit,  are accidentally suppressed in the SM, with respect to the transverse ones~\cite{altro}. It would  therefore be interesting to enhance BSM effects in the longitudinal channel,\footnote{These BSM effects are purely CP even.} by exploiting its interference with the transverse one. The interference terms with the leading SM amplitudes, assuming a fully leptonic final state, reads
\begin{eqnarray}
\hspace{-20pt}&& I_{(00)\otimes(\pm\mp)}^{WZ} = 2 g^2 {\mathcal{A}}_{00}^{\textrm{\tiny{BSM}}_+} \sin\varphi^{\tiny\rm{reco}}_\textrm{\tiny W} \,\sin\hspace{-1pt} \varphi^c_{\textrm{\tiny Z}}\,\,
d_0(\theta_\textrm{\tiny W}) d_0(\theta^c_\textrm{\tiny Z})\times\nonumber\\
\hspace{-20pt}&& \hspace{10pt}\times\Big[g_L^2 [{\mathcal{A}}_{+-}^{\textrm{\tiny{SM}}} d_{+1}(\theta_\textrm{\tiny W}) d_{-1}(\theta^c_\textrm{\tiny Z})
+ {\mathcal{A}}_{-+}^{\textrm{\tiny{SM}}} d_{-1}(\theta_\textrm{\tiny W}) d_{+1}(\theta^c_\textrm{\tiny Z})]\nonumber\\
\hspace{-20pt}&& \hspace{10pt} -\, g_R^2 [{\mathcal{A}}_{+-}^{\textrm{\tiny{SM}}} d_{+1}(\theta_\textrm{\tiny W}) d_{+1}(\theta^c_\textrm{\tiny Z})
+ {\mathcal{A}}_{-+}^{\textrm{\tiny{SM}}} d_{-1}(\theta_\textrm{\tiny W}) d_{-1}(\theta^c_\textrm{\tiny Z})]\Big]\,,\nonumber
\end{eqnarray}
having summed over the $W$ and $Z$ ambiguities.
Differently from the interference terms between the $(\pm\pm)$ and $(\pm\mp)$ transverse channels, the above formula  cancels if integrating over either $\varphi_\textrm{\tiny W}^{\tiny\rm{reco}}$ or $\varphi^c_\textrm{\tiny Z}$, and interference effects can be observed only if the azimuthal decay angles of both gauge bosons are measured. A further subtlety is connected to the polar decay angles. Integrating over the $Z$ polar decay angle $\theta^c_\textrm{\tiny Z}$, leads to
\begin{eqnarray}
\hspace{-25pt}I_{(00)\otimes(\pm\mp)}^{WZ} &=& \frac{\pi}{2\sqrt{2}} g^2 [g_L^2 - g_R^2] {\mathcal{A}}_{00}^{\textrm{\tiny{BSM}}_+}\hspace{-2pt} \sin\varphi^{\tiny\rm{reco}}_\textrm{\tiny W}\hspace{-1pt} \sin\hspace{-1pt} \varphi^c_{\textrm{\tiny Z}}
\times\nonumber\\
\hspace{-25pt}&& \hspace{-10pt}\times d_0(\theta_\textrm{\tiny W})\hspace{-1pt}\Big[{\mathcal{A}}_{+-}^{\textrm{\tiny{SM}}} d_{+1}(\theta_\textrm{\tiny W})
+ {\mathcal{A}}_{-+}^{\textrm{\tiny{SM}}} g_L^2 d_{-1}(\theta_\textrm{\tiny W})\Big]\,,
\label{eq:WZ_long_interf_Z}
\end{eqnarray}
which is suppressed by the small value of $g_L^2 - g_R^2$ due to the almost exclusively axial couplings of the $Z$ boson to the charged leptons ($g_L \simeq - g_R$). 
Integrating over $\theta_\textrm{\tiny W}$ leads instead to no suppression
\begin{gather}
\hspace{-15pt}I_{(00)\otimes(\pm\mp)}^{WZ} = \frac{\pi}{2\sqrt{2}} g^2 {\mathcal{A}}_{00}^{\textrm{\tiny{BSM}}_+}\hspace{-2pt}
 \sin\varphi^{\tiny\rm{reco}}_\textrm{\tiny W}\hspace{-1pt} \sin\hspace{-1pt} \varphi^c_{\textrm{\tiny Z}}\,\,
d_0(\theta^c_\textrm{\tiny Z}) \Big[{\mathcal{A}}_{+-}^{\textrm{\tiny{SM}}} \times \nonumber\\
[g_L^2 d_{-1}(\theta^c_\textrm{\tiny Z})
- g_R^2 d_{+1}(\theta^c_\textrm{\tiny Z})] +\,{\mathcal{A}}_{-+}^{\textrm{\tiny{SM}}} [g_L^2 d_{+1}(\theta^c_\textrm{\tiny Z})
- g_R^2 d_{-1}(\theta^c_\textrm{\tiny Z})]\Big]\,.\nonumber
\label{eq:WZ_long_interf_W}
\end{gather}
Exploiting interference resurrection for new physics in the longitudinal $WZ$ channel is thus particularly challenging in the leptonic $Z$ final state, since it requires the determination of at least three decay angles, namely $\theta^c_\textrm{\tiny Z}$, $\varphi^c_\textrm{\tiny Z}$ and~$\varphi_\textrm{\tiny W}^{\tiny\rm{reco}}$.

\section{Conclusions and Outlook}

Many processes involving electro-weak bosons have dominant SM and BSM (dimension-6 EFT operators) amplitudes with different helicities, hence suppressed interference in inclusive measurements. This can be an important obstacle in the LHC precision program. 
For transverse polarizations this can be understood through simple helicity selection rules that hold  in the high-energy limit, while for the longitudinals it is due to an accidental suppression of the longitudinal SM amplitude.

We have described a method, based on exclusive measurements of azimuthal angular distributions,  that provides enhanced sensitivity to the interference between SM and BSM effects in diboson processes. At the practical level, ambiguities stemming from the $W$-reconstruction procedure, and the impossibility of accessing experimentally the fermion-helicity, singles out a number of processes, and effects, that suit our proposed analysis. In particular, we have estimated the LHC reach  for CP-even modifications of trilinear gauge couplings, using leptonic $W\gamma$ final states, see fig.~\ref{fig:fit}. 
We have verified the robustness of our results with a detailed simulation including detector effects, and assessed the impact of luminosity and systematics. We confirmed that accessing the interference substantially improves the BSM reach. 
The same analysis can be applied to hadronic $W\gamma$ or $WZ$ final states (see also \cite{Azatov:2017kzw}) and access both CP-even and CP-odd effects.

An alternative strategy to resurrect the interference~\cite{Dixon:1993xd} relies on the emission of one extra parton, that turns on same-sign SM and opposite-sign BSM transverse helicity amplitudes. The approach described in this paper is more universally applicable than the latter one, and it does not rely on next-to-leading order parton emission, which is potentially suppressed. Other domains of applicability of our method include the study of longitudinally polarized vector bosons scattering, that appears in this context as one of the most interesting cases because of the very severe accidental suppression of the longitudinal with respect to the transverse and because of the BSM relevance of longitudinal vector boson scattering. However it does not fall in the diboson category we considered in this paper and its study is left to future work. We also leave to future work a complete classification of diboson processes, which one could study with leptonically or hadronically decaying bosons, as well as a fully-differential study using polar, as well as azimuthal distributions. 

\section*{Acknowledgements}

We thank Roberto Franceschini, Thomas Gehrmann, Alex Pomarol and Riccardo Rattazzi for useful discussions.
The work of G.~P. is partly supported by MINECO under Grant CICYT-FEDER-FPA2014-55613-P, FPA2015-64041-C2-1-P, by the Severo Ochoa Excellence Program of MINECO under the grant SO-2012-0234 and by the Generalitat de Catalunya grant 2014-SGR-1450.



\section*{References}

\begin{thebibliography}{99}

\bibitem{Grzadkowski:2010es}
  B.~Grzadkowski, M.~Iskrzynski, M.~Misiak and J.~Rosiek,
  JHEP {\bf 1010} (2010) 085
  [\hhref{1008.4884}].
  
\bibitem{Dixon:1993xd}
L.~J.~Dixon and Y.~Shadmi,
Nucl.\ Phys.\ B {\bf 423} (1994) 3
Erratum: [Nucl.\ Phys.\ B {\bf 452} (1995) 724]
[\hhref{hep-ph/9312363}].

\bibitem{Krauss:2016ely}
  F.~Krauss, S.~Kuttimalai and T.~Plehn,
  Phys.\ Rev.\ D {\bf 95} (2017) no.3,  035024
  [\hhref{1611.00767}].

\bibitem{Alioli:2017jdo}
S.~Alioli, M.~Farina, D.~Pappadopulo and J.~T.~Ruderman,
JHEP {\bf 1707} (2017) 097
[\hhref{1706.03068}].

\bibitem{Farina:2016rws}
M.~Farina, G.~Panico, D.~Pappadopulo, J.~T.~Ruderman, R.~Torre and A.~Wulzer,
Phys.\ Lett.\ B {\bf 772} (2017) 210
[\hhref{1609.08157}].

\bibitem{Hagiwara:1986vm}
K.~Hagiwara, R.~D.~Peccei, D.~Zeppenfeld and K.~Hikasa,
Nucl.\ Phys.\ B {\bf 282} (1987) 253.

\bibitem{Falkowski:2015jaa}
A.~Falkowski, M.~Gonzalez-Alonso, A.~Greljo and D.~Marzocca,
Phys.\ Rev.\ Lett.\ {\bf 116} (2016) no.1, 011801
[\hhref{1508.00581}].

\bibitem{Frye:2015rba}
C.~Frye, M.~Freytsis, J.~Scholtz and M.~J.~Strassler,
JHEP {\bf 1603} (2016) 171
[\hhref{1510.08451}].

\bibitem{Butter:2016cvz}
A.~Butter, O.~J.~P.~ƒEboli, J.~Gonzalez-Fraile, M.~C.~Gonzalez-Garcia, T.~Plehn and M.~Rauch,
JHEP {\bf 1607} (2016) 152
[\hhref{1604.03105}].

\bibitem{Green:2016trm}
D.~R.~Green, P.~Meade and M.~A.~Pleier,
[\hhref{1610.07572}].

\bibitem{Zhang:2016zsp}
  Z.~Zhang,
  Phys.\ Rev.\ Lett.\  {\bf 118} (2017) no.1,  011803
  [\hhref{1610.01618}].
  
\bibitem{Baglio:2017bfe} 
J.~Baglio, S.~Dawson and I.~M.~Lewis,
  \hhref{1708.03332}.

\bibitem{Ellis:2014dva}
J.~Ellis, V.~Sanz and T.~You,
JHEP {\bf 1407} (2014) 036
[\hhref{1404.3667}].

\bibitem{Biekoetter:2014jwa}
A.~Biekoetter, A.~Knochel, M.~Kr\"aŠmer, D.~Liu and F.~Riva,
Phys.\ Rev.\ D {\bf 91} (2015) 055029
[\hhref{1406.7320}].

\bibitem{Azatov:2016sqh}
  A.~Azatov, R.~Contino, C.~S.~Machado and F.~Riva,
  Phys.\ Rev.\ D {\bf 95} (2017) no.6,  065014
  [\hhref{1607.05236}].

\bibitem{Duncan:1985ij}
M.~J.~Duncan, G.~L.~Kane and W.~W.~Repko,
Phys.\ Rev.\ Lett.\ {\bf 55} (1985) 773.

\bibitem{Hagiwara:1989mx}
K.~Hagiwara, J.~Woodside and D.~Zeppenfeld,
Phys.\ Rev.\ D {\bf 41} (1990) 2113.
  
\bibitem{GPFRtalk}
Giuliano Panico, \href{https://indico.cern.ch/event/587148/contributions/2409109/attachments/1393682/2123873/Panico_EWPT_at_LHC.pdf}{talk at ZPW 2017}, 11.1.2017, University of Z\"urich; Francesco Riva, \href{https://indico.cern.ch/event/550509/contributions/2413296/attachments/1396826/2129753/100TeV.pdf}{talk at the 1st FCC Physics Workshop}, 17.1.2017, CERN.  
  
\bibitem{Azatov:2017kzw}
  A.~Azatov, J.~Elias-Miro, Y.~Reyimuaji and E.~Venturini,
 \hhref{1707.08060}.
  
\bibitem{Jacob:1959at}
  M.~Jacob and G.~C.~Wick,
  Annals Phys.\  {\bf 7} (1959) 404.

\bibitem{Murayama:1992gi}
  H.~Murayama, I.~Watanabe and K.~Hagiwara,
  KEK-91-11.
  
\bibitem{Sirunyan:2017tqd}
  A.~M.~Sirunyan {\it et al.} [CMS Collaboration],
  \hhref{1707.00541}.

\bibitem{ATLAS:2017cju}
  The ATLAS collaboration,
  ATLAS-CONF-2017-043.


\bibitem{Alwall:2014hca}
  J.~Alwall {\it et al.},
  JHEP {\bf 1407} (2014) 079
  [\hhref{1405.0301}].
  
  \bibitem{Degrande:2012wf}
  C.~Degrande, N.~Greiner, W.~Kilian, O.~Mattelaer, H.~Mebane, T.~Stelzer, S.~Willenbrock and C.~Zhang,
  Annals Phys.\ {\bf 335} (2013) 21
  [\hhref{1205.4231}].
  
\bibitem{deFavereau:2013fsa}
  J.~de Favereau {\it et al.} [DELPHES 3 Collaboration],
  JHEP {\bf 1402} (2014) 057
  [\hhref{1307.6346}].
  
  \bibitem{Sjostrand:2014zea}
    T.~S\''ojstrand {\it et al.},
    Comput.\ Phys.\ Commun.\  {\bf 191} (2015) 159
    [\hhref{1410.3012}].
    
    \bibitem{Chatrchyan:2013fya}
  S.~Chatrchyan {\it et al.} [CMS Collaboration],
  Phys.\ Rev.\ D {\bf 89} (2014) no.9,  092005
  [arXiv:1308.6832 [hep-ex]].
    
    \bibitem{ATLAS_Photon}
  ATLAS Collaboration,
  ATLAS-PUB-2016-026.
    
\bibitem{Busoni:2014sya}
  G.~Busoni, A.~De Simone, J.~Gramling, E.~Morgante and A.~Riotto,
  JCAP {\bf 1406} (2014) 060
  [\hhref{1402.1275}].

\bibitem{Racco:2015dxa}
  D.~Racco, A.~Wulzer and F.~Zwirner,
  JHEP {\bf 1505} (2015) 009
  [\hhref{1502.04701}].

\bibitem{Contino:2016jqw}
R.~Contino, A.~Falkowski, F.~Goertz, C.~Grojean and F.~Riva,
JHEP {\bf 1607} (2016) 144
[\hhref{1604.06444}].
    
\bibitem{Liu:2016idz}
D.~Liu, A.~Pomarol, R.~Rattazzi and F.~Riva,
JHEP {\bf 1611} (2016) 141
[\hhref{1603.03064}].

\bibitem{altro}
R.~Franceschini, G.~Panico, A.~Pomarol, F.~Riva and A.~Wulzer,
  \hhref{1712.01310}.


\end{thebibliography}
\end{document}